# Comment on "Entanglement and the Thermodynamic Arrow of Time" and Correct Reply on "Comment on "Quantum Solution to the Arrow-of-Time Dilemma"" of David Jennings and Terry Rudolph

Kupervasser Oleg

Scientific Research Computer Center Moscow State University 119992 Moscow, Russia
olegkup@yahoo.com

**Abstract** Recently David Jennings and Terry Rudolph published two papers as reaction on Maccone's paper "Quantum Solution to the Arrow-of-Time Dilemma". In these papers, the authors suppose that second law of thermodynamics is not relevant for quantum systems. Unfortunately, these papers did not get relevant reply from Maccone. The reason of this is following. Both Maccone and the above-mentioned authors use thermodynamic law and thermodynamic-like terminology for non-thermodynamic systems, for example, microscopic system of three qubits. However, big size of a system (quantum or classic) is also not an enough condition for a system to be macroscopic. The macroscopic system must also be chaotic and has small chaotic interaction with its environment/observer resulting in decoherence (decorrelation). We demonstrate that for relevant thermodynamic macroscopic quantum systems no objection appears.

## 1. Introduction

The paper of David Jennings, Terry Rudolph "Entanglement and the Thermodynamic Arrow of Time" is very interesting. However, the *Thermodynamic* Arrow of Time is not applicable for microsystems. It is a nice paper about quantum fluctuation, but not a paper about *Thermodynamic* Arrow of Time. In the Abstract of the paper, "Entanglement and the Thermodynamic Arrow of Time" the authors write: "We examine in detail the case of three qubits, and also propose some simple experimental demonstrations possible with small numbers of qubits." Nevertheless, no thermodynamics is possible for such a microsystem. D. Jennings and T. Rudolph (like Maccone) do not understand that category "*thermodynamic* arrow of time" is correct only for large macrosystems. Using these categories for small fluctuating systems has no physical sense. They also (like Maccone) use incorrect definition of macroscopic *t h e r m o d y n a m i c   e n t r o p y* .

We also give (instead of Maccone) the correct reply to "Comment on "Quantum Solution to the Arrow-of-Time Dilemma"". The *correct* reply is that no contradictions (found in this Comment) appear for macroscopic systems. Only for a microscopic system, such contradictions exist.

However, the concepts "the Thermodynamic Arrow of Time" and "the entropy growth law" is not relevant for such systems. We illustrate this fact by consideration of a quantum chaotic macrosystem and demonstrate that no contradiction (found by David Jennings, Terry Rudolph for a microscopic system) exists for this correct thermodynamical case. It must be mentioned that big size of a system (quantum or classic) is also not an enough condition for a system to be macroscopic. The macroscopic system (considered in Thermodynamics) must also be chaotic (quantum or classic) and has small chaotic interaction with its environment/observer resulting in decoherence (for quantum mechanics) or decorrelation (for classical mechanics). It should be also mentioned that thermodynamic-like terminology is widely and effectively used in quantum mechanics, quantum computers field, and information theory. The big number of the examples can be found in the references of Jennings and Rudolph's paper. The other nice example is Shannon's entropy in information theory. But usually an author (using such a thermodynamic-like terminology) does not consider such a paper as analysis of classical Thermodynamics. Contrarily Jennings and Rudolph "disprove" the second law of Thermodynamics on the basis of the irrelevant microscopic system (in their Comment) and give (also in this Comment) the announcement of their next paper «Entanglement and the *Thermodynamic* Arrow of Time" as a correct consideration and a disproof of the second law. The reason of alignment of thermodynamic time arrows in a quantum mechanics, as well as in the classical mechanics, is small interaction between real chaotic macroscopic systems. This well studied appearance carrying a title «decoherence»[2-3, 17, 24-27]. Its result is not only widely known «entangling» states of systems, but also alignment of thermodynamic time arrows. (The direction of thermodynamic time arrow is defined by a direction of the entropy increase.) The reason of alignment of thermodynamic time arrows is the same, as in the classical Hamilton mechanics - instability of processes with opposite time arrows with respect to small perturbations. These perturbations exist between the observer/environment and observed system (decoherence).

Similar arguments in the case of quantum mechanics have been given in Maccone's paper [4]. However there he formulated, that the similar logic is applicable only in a quantum mechanics. The incorrectness of this conclusion has been shown in our previous papers [1, 5]. The other objection has been formulated in the paper [6]. There are considered small systems with strong fluctuations. Alignment of thermodynamic time arrows does not exist for such small systems. It must be mentioned that both Maccone's replay to this objection and the subsequent paper of objection authors [7] do not explain the true reason of described disagreement. The real solution is very simple. More specifically, the entropy increase law, the concept of thermodynamic time arrows and their alignment are applicable only to nonequilibrium *macroscopic* objects. Violation of these laws for microscopic systems with strong fluctuations is widely known fact. Nevertheless, though the objection[6] is trivial physically, but it is interesting from purely mathematical point of view. It gives good mathematical criterion for macroscopicity of chaotic quantum systems.

## 2. Decoherence for Measurement Process

### 2.1. Reduction of System at Measurement

This part is based on[22, 23]

Let's consider a situation when a measuring device was at the beginning in state $|\alpha_0\rangle$, and the object was in superposition of states $|\psi\rangle = \sum c_i |\psi_i\rangle$, where $|\psi_i\rangle$ are experiment eigenstates. The initial statistical operator is given by expression

$$\rho_0 = |\psi\rangle|\alpha_0\rangle\langle\alpha_0|\langle\psi| \qquad (1)$$

The partial track of this operator which is equal to statistical operator of the system, including only the object, looks like $tr_A(\rho_0) = \sum_n \langle\varphi_n|\rho_0|\varphi_n\rangle$

where $|\varphi_n\rangle$ - any complete set of device eigenstates. Thus,

$$tr_A(\rho_0) = \sum_n |\psi\rangle\langle\varphi_n|\alpha_0\rangle\langle\alpha_0|\varphi_n\rangle\langle\psi| = |\psi\rangle\langle\psi|, \qquad (2)$$

Where the relation $\sum |\varphi_n\rangle\langle\varphi_n| = 1$ and normalization condition for $|\alpha_0\rangle$ are used. We have statistical operator correspondent to object state $|\psi\rangle$. After measuring there is a correlation between device and object states, so the state of full system including device and object is featured by a state vector

$$|\psi\rangle = \sum c_i e^{j\theta_i} |\psi_i\rangle|\alpha_i\rangle \qquad (3)$$

And the statistical operator is given by expression

$$\rho = |\psi\rangle\langle\psi| = \sum c_i c_j^* e^{j(\theta_i - \theta_j)} |\psi_i\rangle|\alpha_i\rangle\langle\alpha_j|\langle\psi_j|. \qquad (4)$$

The partial track of this operator is equal to

$$tr_A(\rho) = \sum_n \langle\varphi_n|\rho|\varphi_n\rangle =$$
$$\sum_{(ij)} c_i c_j^* e^{j(\theta_i - \theta_j)} |\psi_i\rangle \{\sum_n \langle\varphi_n|\alpha_i\rangle\langle\alpha_j|\varphi_n\rangle\} \langle\psi_j| = \qquad (5)$$
$$\sum_{(ij)} c_i c_j^* \delta_{ij} |\psi_i\rangle\langle\psi_j|$$

(Since various states $|\alpha_l\rangle$ of device are orthogonal each other); thus,

$$tr_A(\rho) = \sum_i |c_i|^2 |\psi_i\rangle\langle\psi_i| \qquad (6)$$

We have obtained statistical operator including only the object, featuring probabilities $|c_i|^2$ for object states $|\psi_l\rangle$. So, we come to formulation of the following theorem.

**Theorem 1** (about measuring). If two systems *S* and *A* interact in such a manner that to each state $|\psi_l\rangle$ systems *S* there corresponds a certain state $|\alpha_l\rangle$ of systems *A* the statistical operator $tr_A(\rho)$ over full systems *(S* and *A)* reproduces wave packet reduction for measuring, yielded over system *S,* which before measuring was in a state $|\psi\rangle = \sum_i c_i |\psi_i\rangle$.

Suppose that some subsystem is in mixed state but the full system including this subsystem is in pure state. Such mixed state is named *as improper mixed state.*

### 2.2. The Theorem about Decoherence at Interaction with the Macroscopic Device

This part is based on[18, 84]

Let's consider now that the device is a macroscopic system. It means that each distinguishable configuration of the device (for example, position of its arrow) is not a pure quantum state. It states nothing about a state of each separate arrow molecule. Thus, in the above-stated reasoning the initial state of the device $|\alpha_0\rangle$ should be described by some statistical distribution on microscopic quantum states $|\alpha_{0,s}\rangle$; the initial statistical operator is not given by expression (1), and is equal

$$\rho_0 = \sum_s p_s |\psi\rangle|\alpha_{0,s}\rangle\langle\alpha_{0,s}|\langle\psi| \qquad (7)$$

Each state of the device $|\alpha_{0,s}\rangle$ will interact with each object eigenstate $|\psi_i\rangle$. So, it will be transformed to some other state $|\alpha_{i,s}\rangle$. It is one of the quantum states of set with macroscopic description correspondent to arrow in position i; more precisely we have the formula

$$e^{j\frac{H\tau}{\hbar}}(|\psi\rangle|\alpha_{0,s}\rangle) = e^{j\theta_{i,s}}(|\psi\rangle|\alpha_{i,s}\rangle) \qquad (8)$$

Let's pay attention at appearance of phase factor depending on index *s*. Differences of energies for quantum states $|\alpha_{0,s}\rangle$ should have such values that phases $\theta_{i,s}(mod\ 2\pi)$ after time $\tau$ would be randomly distributed between 0 and $2\pi$.

From formulas (7) and (8) follows that at $|\psi\rangle = \sum_i c_i|\psi_i\rangle$ the statistical operator after measuring will be given by following expression:
$$\rho = \sum_{(s,i,j)} p_s c_i c_j^* e^{j(\theta_{i,s}-\theta_{j,s})}|\psi_i\rangle|\alpha_{i,s}\rangle\langle\alpha_{j,s}|\langle\psi_j| \quad (9)$$

As from (9) the same result (6) can be concluding. So we see that the statistical operator (9) reproduces an operation of reduction applied to given object. It also practically reproduces an operation of reduction applied to device only ("practically" in the sense that it is a question about "macroscopic" observable variable). Such observable variable does not distinguish the different quantum states of the device corresponding to the same macroscopic description, i.e. matrix elements of this observable variable correspondent to states $|\psi_i\rangle|\alpha_{i,s}\rangle$ and $|\psi_j\rangle|\alpha_{j,s}\rangle$ do not depend on $r$ and $s$. Average value of such macroscopic observable variable $A$ is equal to
$$tr(\rho A) = \sum_{(s,i,j)} p_s c_i c_j^* e^{j(\theta_{i,s}-\theta_{j,s})}\langle\alpha_{j,s}|\langle\psi_j|A|\psi_i\rangle|\alpha_{i,s}\rangle$$
$$= \sum_{(i,j)} c_i c_j^* a_{i,j} \sum_s p_s e^{j(\theta_{i,s}-\theta_{j,s})} \quad (10)$$

As phases $\theta_{i,s}$ are distributed randomly, the sum over s are zero at $i \neq j$; hence,
$$tr(\rho A) = \sum_i |c_i|^2 a_{i,i} = tr(\rho' A) \quad (11)$$
Where
$$\rho' = \sum |c_i|^2 p_s|\psi_i\rangle|\alpha_{i,s}\rangle\langle\alpha_{j,s}|\langle\psi_j| \quad (12)$$

We obtain statistical operator which reproduces operation of reduction on the device. If the device arrow is observed in position i, the device state for some $s$ will be $|\alpha_{i,s}\rangle$. The probability to find state $|\alpha_{i,s}\rangle$ is equal to probability of that before measuring its state was $|\alpha_{i,s}\rangle$. Thus, we come to the following theorem.

**Theorem 2. About decoherence of the macroscopic device**. Suppose that the quantum system interacts with the macroscopic device in such a manner that there is a chaotic distribution of states phases of the device. Suppose that $\rho$ is a statistical operator of the device after the measuring, calculated with the help of Schrodinger equations, and $\rho'$ is the statistical operator obtained as a result of reduction application to operator $\rho$. Then it is impossible to yield such experiment with the macroscopic device which would register difference between $\rho$ and $\rho'$. It is the so-called Daneri-Loinger-Prosperi theorem[21]**.**

For a wide class of devices it is proved that the chaotic character in distribution of phases formulated in the theorem 2 really takes place if the device is macroscopic and chaotic with unstable initial state. Indeed, randomness of **phase** appears from randomness of energies (eigenvalues of Hamiltonian) in quantum chaotic systems[8].

It is worth to note that though Eq. (48) is relevant with a split-hair accuracy it is only assumption with respect to (9). There from it is often concluded that the given above proof is FAPP. It means that it is only difficult to measure quantum correlations practically. Actually they continue to exist. Hence, *in principle* they can be measured. It is, however, absolutely untruly. Really, from Poincare's theorem about returns follows that the system will not remain in the mixed state (12), and should return to the initial state (7). It is the result of the very small corrections (quantum correlation) which are not included to (12). Nevertheless, the system featured here $|\alpha_{i,s}\rangle$ corresponds *to the introspection* case, and consequently, it is not capable to observe experimentally these returns *in principle (*as it was shown above in resolution of Poincare and Loshmidt paradoxes). Hence, effects of these small corrections exist only on paper in the coordinate time of ideal dynamics, but it cannot be observed *experimentally* with respect to thermodynamic time arrow of observable dynamics of the macroscopic device. So, we can conclude that Daneri-Loinger-Prosperi theorem actually results in a complete resolution (not only FAPP!) of the reduction paradox *in principle*. It proves impossibility to distinguish *experimentally* the complete and incomplete reduction.

The logic produced here strongly reminds Maccone's paper[4]. It is not surprising. Indeed, the pass from (7) to (12) corresponds to increasing of microstates number and entropy growth. And the pass from (12) in (7) corresponds to the entropy decrease. Accordingly, our statement about experimental unobservability to remainder quantum correlation is equivalent to the statement about unobservability of the entropy decrease. And it is proved by the similar methods, as in [4]. The objection [6] was made against this paper. Unfortunately, Maccone could not give the reasonable reply [28] to this objection. Here we will try to do it ourselves.

Let's define here necessary conditions.

Suppose $A$ is our device, and $C$ is the measured quantum system.

The first value, the mutual entropy $S(A:C)$ is the coarsened entropy of ensemble (received by separation on two subsystems) excluding the ensemble entropy. As the second excluding term is constant, so $S(A:C)$ describes well the behavior of macroentropy in time:
$$S(A:C) = S(\rho_A) + S(\rho_C) - S(\rho_{AC})$$
where $S = -tr(\rho\ln\rho)$

The second value $I(A:C)$ is the classical mutual information. It defines which maximum information about measured system $(F_j)$ we can receive from indication of instrument $(E_i)$. The more correlation exists between systems, the more information about measured system we can receive:
$I(A:C) = \max_{E_i \otimes F_j} H(E_i:F_j)$,
where
$H(E_i:F_j) = \sum_{ij} P_{ij}\log P_{ij} - \sum_i p_i\log p_i - \sum_i q_i\log q_i$,
$P_{ij} = Tr[E_i \otimes F_j \rho_{AC}]$, $p_i = \sum_i P_{ij}$ and $q_j = \sum_i P_{ij}$ - given POVMs (Positive Operator Valued Measure) $E_i$ and $F_j$ for A and C, respectively.

Maccone[4] proves an inequality
$$S(A:C) \geq I(A:C) \quad (13)$$

He concludes from it that entropy decrease results in reduction of the information (memory) about the system $A + C$ and $C$.

But (13) contains an inequality. Correspondingly in[6] an example of the quantum system of three qubits is supplied. For this system the mutual entropy decrease is accompanied by mutual information increases. It does not contradict to (13) because mutual entropy is only up boundary for mutual information there.

Let's look what happens in our case of the macroscopic device and the measured quantum system

Before measurement (7)

$$S(A:C) = -\sum_s p_s \log p_s + 0 + \sum_s p_s \log p_s = 0$$

$E_i$-corresponds to the set $|\alpha_{0,s}\rangle$, $F_j$- $|\psi\rangle$

$$I(A:C) = -\sum_s p_s \log p_s + 0 + \sum_s p_s \log p_s = 0 = S(A:C)$$

In the end of measurement from (12)

$$S(A:C) = -\sum_s |c_i|^2 \log |c_i|^2 - \sum_{s,i} |c_i|^2 p_s \log |c_i|^2 p_s$$
$$+ \sum_{s,i} |c_i|^2 p_s \log |c_i|^2 p_s$$
$$= -\sum_s |c_i|^2 \log |c_i|^2$$

$E_i$-corresponds to the set $|\alpha_{i,s}\rangle$, $F_j$- $|\psi_j\rangle$

$$I(A:C) = -\sum_s |c_i|^2 \log |c_i|^2 - \sum_{s,i} |c_i|^2 p_s \log |c_i|^2 p_s$$
$$+ \sum_{s,i} |c_i|^2 p_s \log |c_i|^2 p_s =$$
$$-\sum_s |c_i|^2 \log |c_i|^2 = S(A:C)$$

Thus, our case corresponds to

$$S(A:C) = I(A:C) \qquad (14)$$

in (13). No problems exist for our case. It is not surprising -- the equality case in (13) corresponds to macroscopic chaotic system. The system supplied by the objection [6] is not macroscopic. It demonstrates the widely known fact that such *thermodynamic* concepts as the thermodynamic time arrows, the entropy increase and the measurement device concern to macroscopic chaotic systems. Both the paper [6] and the subsequent paper [7] describe not thermodynamic time arrows but, mainly, strongly fluctuating small systems. No thermodynamics is possible for such small systems as three cubits. The useful outcome of these papers is equality (14). It can be used as a *measure for macroscopicity* of chaotic quantum systems. On the other hand, the difference between mutual information and mutual entropy can be a criterion of fluctuations value.

## 3. Conclusions

D. Jennings and T. Rudolph (like Maccone) use category "thermodynamic arrow of time" for non-macroscopic systems, for example, small fluctuating quantum systems. As a result, they get objections with the second law of thermodynamics. We demonstrate that for relevant macroscopic quantum thermodynamical systems no objection appears.

## BIBLIOGRAPHY

[1] Oleg Kupervasser, Hrvoje Nikolic and Vinko Zlatic, "The Universal Arrow of Time", Foundations of Physics, May 2012, Online first, http://www.springerlink.com/content/v4h2535hh14uh084/ arXiv:1011.4173

[2] M. Schlosshauer, "Decoherence and the Quantum-to-Classical Transition", Springer, 2007.

[3] Zurek W.H., "Decoherence, einselection, and the quantum origins of the classical", Reviews of modern physics, Vol. 75, no. 3, 2003.

[4] Maccone L., "Quantum solution to the arrow-of-time dilemma", Phys.Rev.Lett., vol. 103, p. 080401, 2009.

[5] Oleg Kupervasser, Dimitri Laikov, "Comment on "Quantum Solution to the Arrow-of-Time Dilemma" of L. Maccone", Online Available: arXiv:0911.2610.

[6] D. Jennings, T. Rudolph, "Comment on "Quantum Solution to the Arrow-of-Time Dilemma" of L. Maccone", Phys. Rev. Lett., Vol. 104, p. 148901, 2010.

[7] D. Jennings, T. Rudolph, "Entanglement and the Thermodynamic Arrow of Time", Phys. Rev. E, Vol. 81, p. 061130, 2010.

[8] Stockmann "Quantum Chaos", Cambridge University Press, 2000.

[9] Stanford encyclopedia of Philosophy: Many-Worlds Interpretation of Quantum Mechanics, Online Available: http://plato.stanford.edu/entries/qm-manyworlds/

[10] O. Kupervasser, Online Available: arXiv:0911.2076.

[11] O. Kupervasser, D. Laikov, arXiv:0911.2610

[12] O. Kupervasser, Online Available: arXiv:nlin/0508025

[13] O. Kupervasser, Online Available: arXiv:nlin/0407033

[14] Ilya Prigogine, "From being to becoming: time and complexity in the physical sciences", W.H. Freeman, San Francisco, 1980.

[15] Karl Blum, "Density Matrix Theory and Applications", Plenum Press, New York, 1981

[16] G. C. Ghirardi, A. Rimini and T. Weber, "A Model for a Unified Quantum Description of Macroscopic and Microscopic Systems. Quantum Probability and Applications", eds L. Accardi et al., Springer, Berlin,1985.

[17] Wheeler, J.A.; Zurek, W.H, "Quantum Theory and Measurement", Princeton University Press, Princeton, N.J, 1983

[18] Klimontovich, L., "Statistical Physics", Harwood, New York, 1986

[19] Jonathon Friedman et al., "Quantum superposition of distinct macroscopic states", Nature, Vol. 406, pp. 43-46, 2000.

[20] Alexey Nikulov, "Comment on "Probing Noise in Flux Qubits via Macroscopic Resonant Tunneling", Online Available: arXiv:0903.3575v1

[21] Daneri A., Loinger A., Prosperi G. M., "Quantum theory of measurement and ergodicity conditions", Nuclear Phys., Vol. 33, pp.297-319, 1962.

[22] Anthony Sudbery, "Quantum Mechanics and the Particles of Nature: An Outline for Mathematicians", Cambridge University Press, New York, 1986

[23] J. von Neumann,"Mathematische Grundlagen der Quantemechanik", Springer, Berlin, 1932

[24] H.D. Zeh, "The Physical Basis of the Direction of Time", Springer, Heidelberg, 2007.

[25] H.D. Zeh, Entropy, Vol. 7, p. 199, 2005.

[26] H.D. Zeh, Entropy, Vol. 8, p. 44. 2006.

[27] Erich Joos , H. Dieter Zeh, Claus Kiefer, Domenico J. W. Giulini, Joachim Kupsch, Ion-Olimpiu Stamatescu, "Decoherence and the Appearance of a Classical World in Quantum Theory", Springer, 2003

[28] Maccone L., "A quantum solution to the arrow-of-time dilemma: reply" , Online Available: arXiv:0912.5394

[29] Avshalom Elitzur , Vaidman L. , "Quantum mechanical interaction – free measurement", Found of Phys., Vol. 23, pp. 987-997, 1993

[30] Albert, D. Z, "Quantum Mechanics and Experience". Harvard University Press, Cambridge, 1992

[31] John Byron Manchak, "Self-Measurement and the Uncertainty Relations, Department of Logic and Philosophy of Science", University of California. Online Available: http://philpapers.org/rec/MANSAT

[32] Rudolf Peierls, "Surprises in theoretical physics", Princeton University Press, Princeton,N.J.,1979